\documentstyle[12pt,epsfig]{article}

\begin{document}
\newcommand{\gsim}{ \mathop{}_{\textstyle \sim}^{\textstyle >} }
\newcommand{\lsim}{ \mathop{}_{\textstyle \sim}^{\textstyle <} }
\newcommand {\be}{\begin{equation}}
\newcommand {\ee}{\end{equation}}
\newcommand {\ba}{\begin{eqnarray}}
\newcommand {\ea}{\end{eqnarray}}
\newcommand {\bea}{\begin{array}}
\newcommand {\cl}{\centerline}
\renewcommand {\thefootnote}{\fnsymbol{footnote}}
\newcommand {\eea}{\end{array}}
\vskip .5cm

\baselineskip 0.65 cm

\def \a'{\alpha'}
\baselineskip 0.65 cm
\begin{flushright}
IPM/P-2005/057\\
IZTECH-P2005-07\\
hep-ph/0508236 \\
\today
\end{flushright}
\begin{center}
{\Large{\bf Can Measurements of Electric Dipole Moments Determine
the Seesaw Parameters?} } {\vskip 0.5 cm}

{\bf  Durmu{\c s} A. Demir$^a$   and Yasaman Farzan$^b$
\\}{\vskip 0.5 cm } $^a$ Department of Physics,
Izmir Institute of Technology, Izmir, TR 35430, Turkey \\
 $^b$
Institute for Studies in Theoretical Physics and Mathematics (IPM)\\
P.O. Box 19395-5531, Tehran, Iran\\
\end{center}
\begin{abstract}
In the context of the supersymmetrized seesaw mechanism embedded
in the Minimal Supersymmetric Standard Model (MSSM), complex
neutrino Yukawa couplings can induce Electric Dipole Moments
(EDMs) for the charged leptons, providing an additional route to
seesaw parameters. However, the complex neutrino Yukawa matrix is
not the only possible source of CP violation. Even in the
framework of Constrained  MSSM  (CMSSM), there are additional
sources, usually attributed to the phases of the trilinear soft
supersymmetry breaking couplings and the mu-term, which contribute
not only to the electron EDM but also to the  EDMs of neutron and
heavy nuclei. In this work, by combining bounds on various EDMs,
we analyze how the sources of CP violation can be discriminated by
the present and planned EDM experiments.
\end{abstract}
\section{Introduction}

The atmospheric and solar neutrino data \cite{solar} as well as
the KamLAND \cite{kamland} and K2K \cite{k2k} results provide
strong evidence for nonzero neutrino mass. On the other hand, from
kinematical studies \cite{mainz} and cosmological observations
\cite{cosmo}, the neutrinos are known to be much lighter than the
other fermions. There are several models that generate tiny yet
nonzero masses for neutrinos (see, $e.g.$ \cite{smirnov}) among
which the seesaw mechanism \cite{seesaw} is arguably the most
popular one. This mechanism introduces three Standard Model (SM)
singlet neutrinos with masses, $M_N$, which lie far above the
electroweak scale. It has been shown that for $M_i>10^9$ GeV,
decays of the right-handed neutrinos in the early Universe can
explain the baryon asymmetry of the universe \cite{leptogenesis}.
In addition to this, $M_N$ lies at intermediate scales which are
already marked by other phenomena including supersymmetry breaking
scale, gauge coupling unification scale and the Peccei-Quinn
scale. This rough convergence of scales of seemingly  distinct
phenomena might be related to their common or correlated origin
dictated by first principles stemming, possibly, from
superstrings. For probing physics at ultra high energies which are
obviously beyond the reach of any man-made accelerator in
foreseeable future, it is necessary to analyze and determine the
effects of right-handed neutrinos on the low-energy observables.

The Minimal Supersymmetric Standard Model (MSSM), a direct
supersymmetrization of the SM using a minimal number of extra
fields, solves the gauge hierarchy problem; moreover, it provides
a natural candidate for cold dark matter in the universe. For
explaining the neutrino data within the seesaw scheme, the MSSM
spectrum should be enlarged by right-handed neutrino
supermultiplets. The resulting model, which we hereafter call
MSSM-RN, is described by the superpotential
 \be
W=Y_\ell^{i j} \epsilon_{\alpha \beta}H_d^\alpha E_{i} L_j^\beta-
Y_\nu^{i j} \epsilon_{\alpha \beta}H_u^\alpha N_{i}
L_j^\beta+\frac{1}{2}M_{ij} N_{i} N_{j}-\mu \epsilon_{\alpha
\beta} H_d^\alpha H_u ^\beta, \ee whose quark sector, not shown
here, is the same as in the MSSM. Here $\alpha, \beta$ are SU(2)
indices, $i,j$ are generation indices, $L_{j \beta}$ consist of
lepton doublets $(\nu_{j L}, \ell^-_{jL})_\beta$, and $E_i$
contain left-handed anti-leptons $\ell^+_{iL}$. The superfields
$N_i$ contain anti right-handed neutrinos. Without loss of
generality, one can rotate and rephase the fields to make Yukawa
couplings of charged leptons ($Y_\ell$) as well as the mass matrix
of the right-handed neutrinos ($M_{ij}$)  real diagonal. In the
calculations below, we will use this basis.

In general, the soft supersymmetry-breaking terms (the
mass-squared matrices and trilinear couplings of the sfermions)
can possess flavor-changing entries which facilitate a number of
flavor-changing neutral current processes in the hadron and lepton
sectors. The existing experimental data thus put stringent bounds
on flavor-changing entries of the soft terms. For instance,
flavor-changing entries of the soft terms in the lepton sector can
result in sizeable $\mu \to e \gamma$, $\tau \to e \gamma$ and
$\tau \to \mu \gamma$.
This motivates us to go to the mSUGRA \cite{msugra} or constrained
MSSM framework where soft terms of a given type unify at the scale
of gauge coupling unification. In other words, at the GUT scale,
we take \ba {\cal L}_{soft} &=&
-m_0^2(\tilde{L}_{i}^\dagger\tilde{L}_{i}+ \tilde{E}_{i}^\dagger
\tilde{E}_{i}+ \tilde{N}_{i}^\dagger \tilde{N}_{i}+H_d^\dagger
H_d+ H_u^\dagger H_u \label{soft}) \cr &-&
 \frac{1}{2} m_{1/2}(\tilde{B} \tilde{B}+
\tilde{W} \tilde{W}+\tilde{g} \tilde{g}+{\rm H.c.}) \cr &-& (
\frac{1}{2}\epsilon_{\alpha \beta} b_H\mu H_d^\alpha H_u^\beta
+{\rm H.c.})-( A_\ell^{ij}\epsilon_{\alpha \beta}
H_d^\alpha\tilde{E}_{i} \tilde{L}_{j}^\beta-
A_\nu^{ij}\epsilon_{\alpha\beta} H_u^\alpha \tilde{N}_{i}
\tilde{L}_{j}^\beta +{\rm H.c.})\cr &-& (\frac{1}{2} B_\nu M_i
\tilde{N}^i\tilde{N}^i+{\rm H. c.}).\label{soft} \ea  Here
$A_\ell=a_0 Y_\ell$ and $A_\nu=a_0 Y_\nu$. The last term is the
lepton number violating neutrino bilinear soft term which is
called the neutrino $B$-term.

As first has been shown in \cite{masiero}, at lower scales, the
Lepton Flavor Violating (LFV) Yukawa coupling $Y_\nu$ will induce
LFV contributions to the soft masses of the left-handed sleptons.
Consequently, the strong bounds on LFV rare decays can be
translated into bounds on the seesaw parameters. In Sec. 4, we
will discuss these bounds in detail. If we assume that the soft
terms are of the form (\ref{soft}) \footnote{In practice,
confirmation of this assumption is not possible. However, this
assumption will be refuted if we find that the flavor mixing in
the right-handed sector is comparable to that in the left-handed
sector.} and $Y_\nu$ is the only source of LFV then mass-squares
of left-handed sleptons can be considered as another source of
information on the seesaw parameters. It is shown in Ref.
\cite{sacha} that, by knowing all the entries of the mass matrices
of neutrinos and left-handed sleptons (both their norms and
phases), we can extract all the seesaw parameters. However, such a
possibility at the moment does not seem to be achievable. As a
result, one has to resort to finding new sources of information on
the seesaw parameters.

In general, the neutrino Yukawa coupling, $Y_\nu$, can possess
CP--odd phases, and thus  induces  electric dipole moments (EDM)
for charged leptons \cite{hisano,peskin}. It has already been
suggested to extract seesaw parameters from the electron EDM,
$d_e$ \cite{suggested}. However, for deriving any information from
$d_e$ we must be aware of other sources of CP violation that can
give a significant contribution to $d_e$. In the model we are
using, there are three extra sources of CP violation in the
leptonic sector: the physical phases of the $\mu$ parameter,  the
universal trilinear coupling $a_0$ and the neutrino
$B$-term\footnote{In fact, apart from the phases of $a_0$ and
$\mu$ there are two more sources of CP-violation: the phase of the
CKM matrix and the QCD theta term. The contribution of the former
to EDMs  of charged leptons is negligible \cite{ckm}. For the
latter we assume that there is a mechanism like the Peccei-Quinn
mechanism that suppresses the CP-odd topological term in the  QCD
Lagrangian.}.  As first has been shown in \cite{yasaman}, the
phase of the neutrino $B$-term can induce a contribution to $d_e$.
In this paper, for simplicity, we will set $B_\nu=0$.
 The phases of $a_0$ and $\mu$ can result in comparable electric
dipole moments for the electron, neutron and mercury. More
precisely, they induce $d_e\sim (m_e/m_d) d_d\sim (m_e/m_u)d_u
\sim e(m_e/m_d) \tilde{d}_d\sim e (m_e/m_u)\tilde{d}_u$, where
$\tilde{d}_u$ and $\tilde{d}_{d}$  respectively are the chromo
electric dipole moments (CEDM) of up and down quarks which
contribute to the EDMs of mercury ($d_{Hg}$) and deuteron ($d_D$).
In principle, the phases of $a_0$ and $\mu$ can induce $d_D$ which
may be detectable in future searches \cite{Dexp}.
 On the other hand, as shown in the appendix, the
quark EDMs and CEDMs induced by the phases of $Y_\nu$ are too
small to be detectable in near future. Therefore, if complex
$Y_\nu$ is the only source of CP violation, we expect $d_D$ to be
too small to be detectable in the near future ($d_D$ is measured
with ionized deuteron which is depleted from electrons). Based on
these observations we raise the following question: Considering
the limited accuracy of the experiments, is it possible to discern
the source of the CP violation? The present paper addresses this
very question.

This paper is organized as follows. In Sec. 2, we show that there
is a ``novel" contribution to $d_\ell$ which is proportional to
$m_{1/2}$, and it results from the renormalization group running
of the trilinear couplings. As will be demonstrated in the text,
the new contribution can dominate over those previously discussed
in the literature. In Sec. 3, we first review the experimental
bounds on the EDMs. We then review how observable EDMs of neutron
and different nuclei are related to the  EDMs and CEDMs of the
quarks. In Sec. 4, we represent our numerical results and analyze
the prospects of identifying the source of CP violation.
Conclusions are given in Sec. 5.

\section{Contribution of $Y_\nu$ to EDMs}
In this section, we review the effects of complex $Y_\nu$ on the
charged lepton EDMs which has been previously calculated in the
literature. We also discuss a new effect which has been so far
overlooked. In the end, we point out an unexpected suppression
that occurs when we insert realistic values for the mSUGRA
parameters. Throughout this section we will assume that complex
$Y_\nu$ is the only source of CP-violation.

\begin{center}
\begin{figure}
\hskip 3 cm \psfig{figure=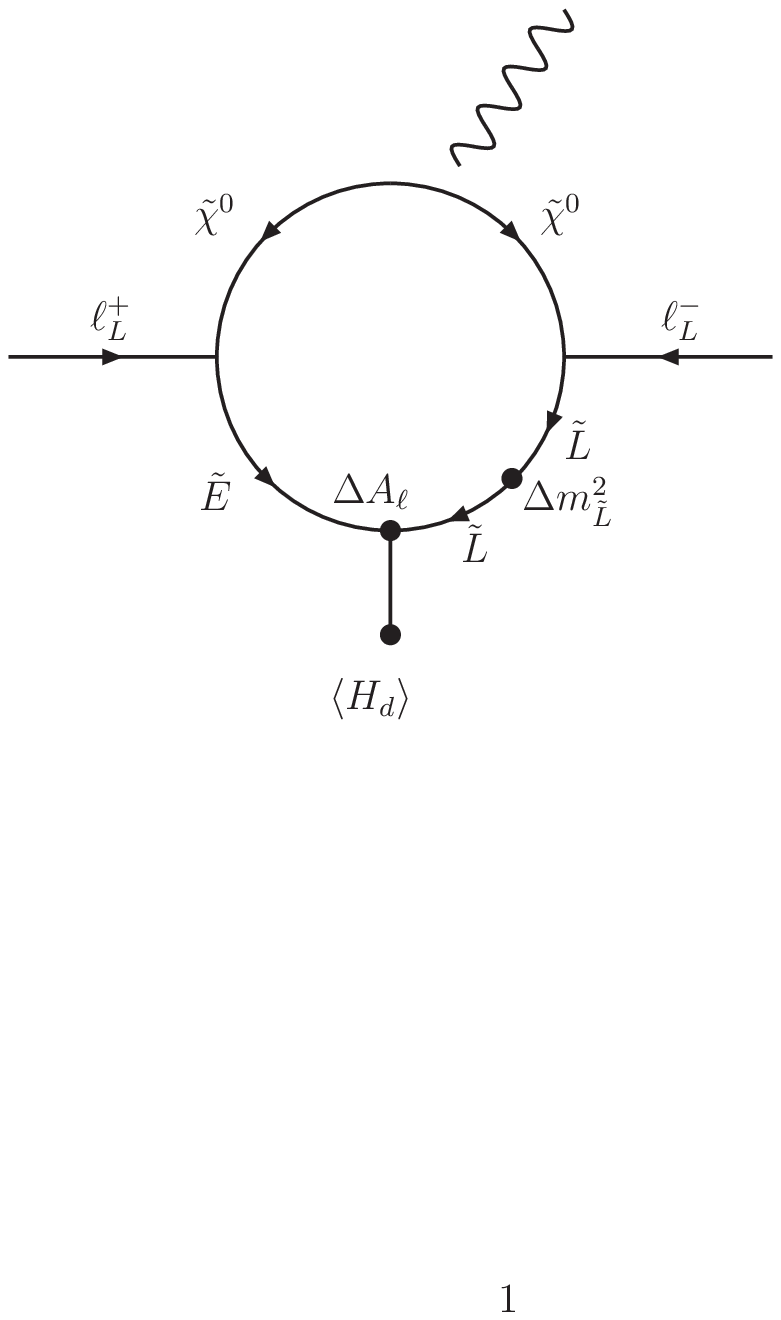,bb=140 233 395 490, clip=true,
height=4 in} \caption{A contribution to the charged lepton dipole
moments.}
 \label{figEDM}
\end{figure}
\end{center}

As it is shown in \cite{peskin}, inserting  LFV radiative
corrections to $A_\ell$ and $ m_{\tilde{L}}^2$  in the diagram
shown in Fig. (\ref{figEDM}), we obtain a contribution to the EDM
of the corresponding charged lepton. By inserting one-loop lepton
flavor violating corrections to $A_\ell $ and $ m_{\tilde{L}}^2$,
we obtain \ba \vec{d}_i^{\ (1)} &=& (-e)\eta_{d_e}m_{\ell i}{2
\alpha \over (4 \pi)^5}\sum_a \sum_{k,j,m} \left({V_{1a} \over
c_w} \right)\left({V_{1a} \over c_w}+{V_{2a} \over s_w}\right){
a_0 m_a \over |m_a|^6}g({m^2_{\tilde{L}}\over
m_a^2},{m^2_{\tilde{E}}\over m_a^2})\cr &\cdot& {\rm
Im}[(Y_\nu^{ki})^*Y_\nu^{kj} (Y_\nu^{mj})^* Y_\nu^{mi}]\cdot
(-2m_0^2 {\rm Log} {M_{GUT}^2 \over M_k^2})\vec{S}  , \label{dea0}
\ea where $\vec{S}$ is the spin of the lepton, $V$ is the mixing
matrix of the neutralinos, $m_a$ are the masses of the neutralinos
and \ba g(x_L,x_E)&=& { 1 \over 2 (x_E-x_L)^2} \left( {1-x_L^2+2
x_L {\rm Log}x_L \over (1-x_L)^3}-{1-x_E^2+2 x_E{\rm Log}x_E \over
(1-x_E)^3} \right) \cr &+& {1 \over 2 (x_E-x_L)}\left({
5-4x_L-x_L^2+2(1+2x_L){\rm Log} x_L \over (1-x_L)^4}\right) . \ea

The  main contribution to the diagram shown in Fig. (\ref{figEDM})
comes from the momenta around the supersymmetry breaking scale
($M_{susy}$); as a result we have to insert the values of $\Delta
A_\ell$ and $\Delta m_{\tilde{L}}^2$ at $M_{susy}$ by taking into
account the effects of running of the effective operators from the
scale that the right-handed neutrinos decouple down to $M_{susy}$.
It can be shown that the LFV corrections to the slepton masses
remain unchanged between the two scales. However, lowering energy
from the right-handed neutrino scale down to $M_{susy}$, $\Delta
A_\ell$ changes significantly. Here, the main effect comes from
the gauge interaction and we can practically neglect the effects
of $Y_\ell$ on the running. The factor $\eta_{d_e}\simeq 1.5$ in
Eq. (\ref{dea0}) takes care of the running of $\Delta A_\ell$.

\begin{figure}
\psfig{figure=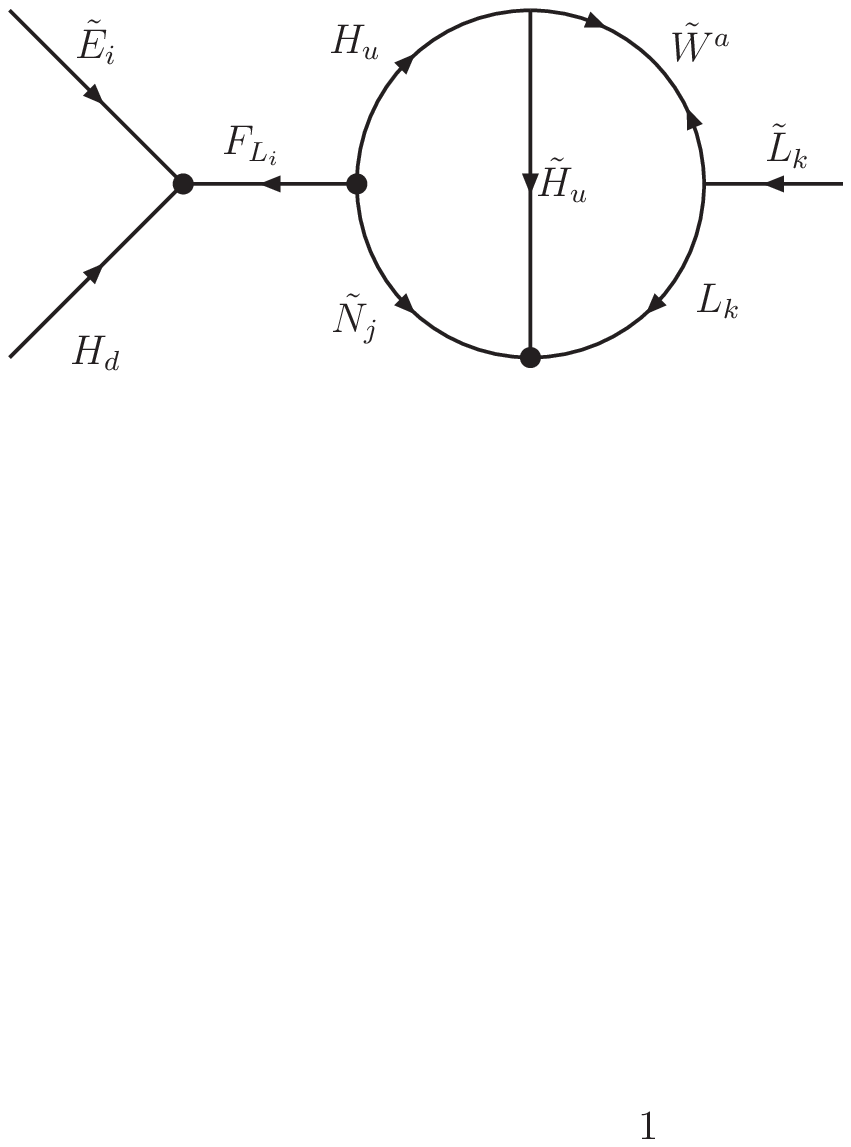,bb=108 306 370 448, clip=true,
height=2.5 in} \caption{The two-loop correction to $A_\ell$ given
by $m_{1/2}$. Vertices marked with circles are Yukawa vertices and
the rest are gauge vertices. $F_{L_i}$ is the auxiliary field
associated with $L_i$.}
 \label{a0m1bar2}
\end{figure}

\begin{figure}
\psfig{figure=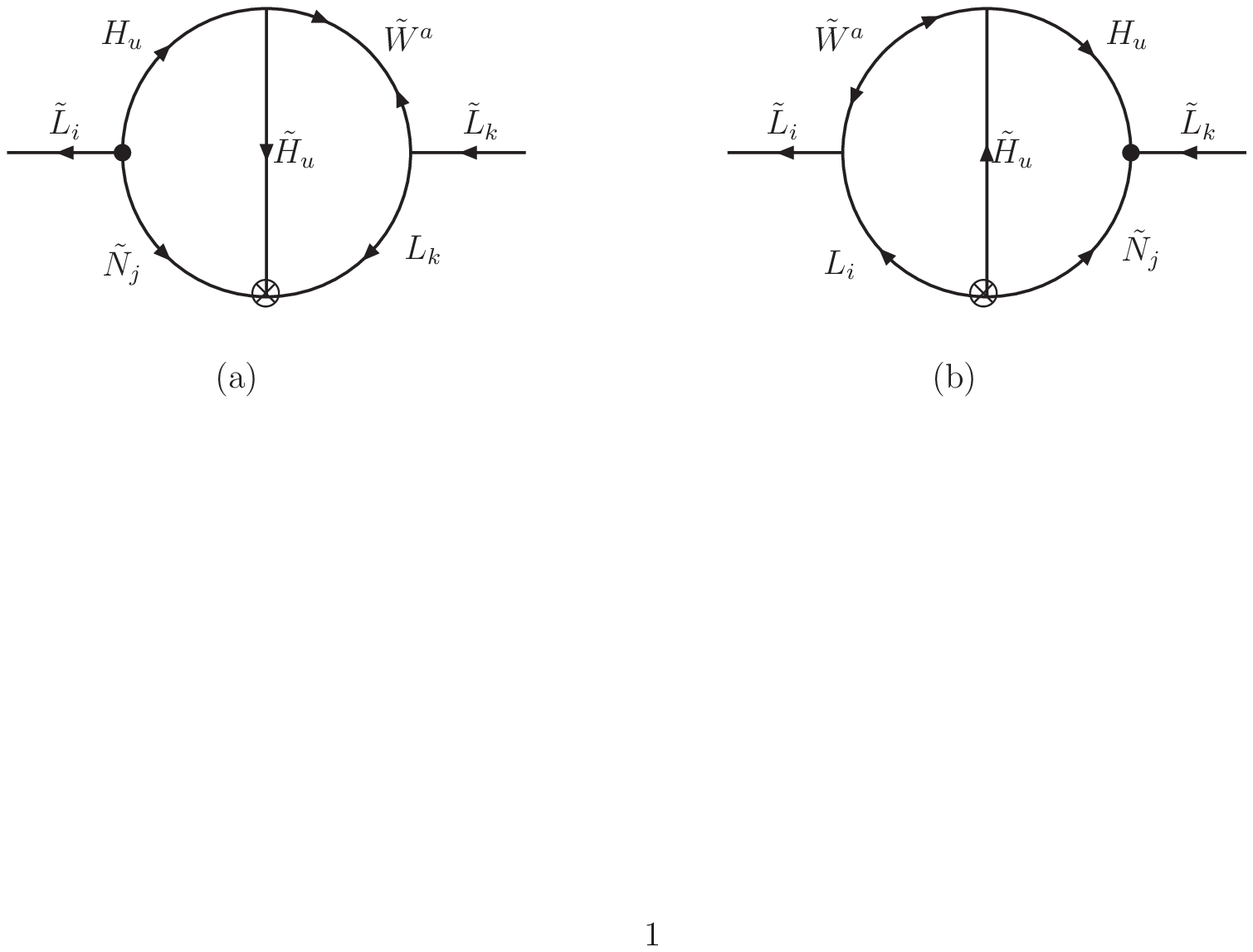,bb=65 274 530 448, clip=true,
height=2.5 in} \caption{The two-loop corrections to
$m_{\tilde{L}}$ given by $m_{1/2}$. Vertices marked with $\otimes$
and circles are Yukawa  vertices and $A$-terms, respectively. The
rest are gauge vertices.}
 \label{MLm1bar2}
\end{figure}

Now, let us discuss the running of the relevant parameters from
the GUT scale down to the right-handed neutrino scale. Let us take
$M_{GUT}=2 \times 10^{16}$ GeV and $M_N\sim Y_\nu^2 \langle H_u
\rangle^2/m_\nu$. For $Y_\nu\sim 1$ and $m_\nu \sim 0.1$ eV, we
find $M_N\sim 10^{14}$ GeV so we expect that the running of
parameters from the GUT scale down to the right-handed neutrino
mass scale to be suppressed by ${\rm Log} (M_N^2/M_{GUT}^2)/(16
\pi^2)\sim 0.1$. Thus, we can practically neglect the running of
the gauge and Yukawa couplings as well as the gaugino and
right-handed neutrino masses in this range. But there is a
subtlety to be noted here. Although the dominant terms of both
$\Delta A_\ell$ and $\Delta m_{\tilde{L}}^2$ are enhanced by a
large log factor ${\rm Log} {M_{GUT}^2 \over M_k^2}$, the effect
in Eq. (\ref{dea0}), which is given by ${\rm Im}(\Delta A_\ell
\Delta m_{\tilde{L}}^2 )$, contains only one factor of ${\rm Log}
{M_{GUT}^2 \over M_k^2}$. This is because the leading-log parts of
$\Delta A_\ell$ and $\Delta m_{\tilde{L}}^2$  have the same flavor
structure $\sum_k (Y_\nu^{ki})^* {\rm Log} (M_{GUT}^2/M_k^2)
Y_\nu^{kj}$, and thus, ${\rm Im} \left[ (\Delta
A_\ell)_{leading-log} (\Delta
m_{\tilde{L}}^2)_{leading-log}\right]=0 $ and the dominant
contribution to ${\rm Im}[\Delta A_\ell \Delta m_{\tilde{L}}^2 ]$
comes from ${\rm Im} \left[ (\Delta A_\ell)_{leading-log} (\Delta
m_{\tilde{L}}^2)_{finite}\right]$ and ${\rm Im} \left[ (\Delta
A_\ell)_{finite} (\Delta m_{\tilde{L}}^2)_{leading-log}\right]$
which contain only one large log factor. If there is a two-loop
contribution to the $A_\ell$ term or mass matrix of the
left-handed sleptons $\left[(\Delta A_\ell)_{2-loop}\right.$ or
$\left. (\Delta m_{\tilde{L}}^2)_{2-loop}\right]$  with two
large-log factors, ${\rm Im}[\Delta A_\ell^{2-loop} (\Delta
m_{\tilde{L}}^2)^{1-loop}_{L-L})]$ and ${\rm Im}[(\Delta
A_\ell)^{1{\rm -}loop}_{L-L} (\Delta m_{\tilde{L}}^2)^{2-loop})]$
(here $L-L$ indices denote leading-log contributions) can be
comparable to ${\rm Im}\left[(\Delta A_\ell)_{L-L}^{1-loop}(\Delta
m_{\tilde{L}}^2)_{finite}^{1-loop}\right].$ Consequently,
inserting the 2-loop correction to $A_\ell$ and 1-loop correction
to $\Delta m_{\tilde{L}}^2$ (or vice-versa) in the diagram shown
in Fig (\ref{figEDM}), we get an effect comparable to (or dominant
over) Eq. (\ref{dea0}). The diagrams shown in Figs.
(\ref{a0m1bar2}) and (\ref{MLm1bar2}) give the dominant two-loop
corrections to   ($\Delta A_\ell$) and ($\Delta m_{\tilde{L}}^2$),
respectively. The leading-log parts of the diagrams are
\footnote{Note that there are similar diagrams with $\tilde{B}$
replacing $\tilde{W}$ in the loops. The effects of the latter is
less than 20\% of the ones we are considering here. Such a
precision is beyond the scope of this paper.}\be (\Delta
A_\ell)_{ik}=\frac{3}{2}m_{1/2} \frac{g^2} {(4\pi)^4}\sum_j
Y_\ell^i(Y_\nu^{ji})^*Y_\nu^{jk} \left( {\rm
Log}\frac{M_{GUT}^2}{M_j^2}\right)^2 \ee and \be (\Delta
m_{\tilde{L}}^2)_{ik}=3 m_{1/2}a_0 \frac{g^2} {(4\pi)^4} \sum_j
(Y_\nu^{ji})^*Y_\nu^{jk} \left( {\rm
Log}\frac{M_{GUT}^2}{M_j^2}\right)^2. \ee Inserting these diagrams
in the diagram shown in Fig (\ref{figEDM}) we arrive at the
following result \ba \vec{d}_i^{\ (2)} &=& (-e)\eta_{d_e} m_{\ell
i}{-2 \alpha \over (4 \pi)^7}\frac{3g^2}{2}\sum_a \sum_{k,j,m}
\left({V_{1a} \over c_w} \right)\left({V_{1a} \over c_w}+{V_{2a}
\over s_w}\right){ m_{1/2} m_a \over
|m_a|^6}g({m^2_{\tilde{L}}\over m_a^2},{m^2_{\tilde{E}}\over
m_a^2})\cr &\cdot& {\rm Im}[(Y_\nu^{ki})^*Y_\nu^{kj}
(Y_\nu^{mj})^* Y_\nu^{mi}]\cdot (3m_0^2-a_0^2) \left({\rm Log}
{M_{GUT}^2 \over M_k^2}\right)^2{\rm Log} { M_{GUT}^2 \over
M_m^2}\vec{S} . \label{dem1bar2} \ea This effect had been
overlooked in the literature.

 Finally, as discussed in
\cite{peskin}, in large $\tan\beta$ domain the dominant
contribution takes the following form: \ba \vec{d}_i^{\ (3)} &=& e
{8 \alpha \over (4 \pi)^7}\sum_a\left({V_{1a} \over
c_w}\right)\left( {V_{1a} \over c_w}+{V_{2 a} \over s_w}\right)
{\mu m_{\ell i} m_a \over |m_a|^8 v^2} {\tan \beta \over \cos^2
\beta}(9m_0^4+9a_0^2m_0^2+2a_0^4)h(\frac{m_{\tilde{L}}^2}{m_a^2},\frac{m_{\tilde{E}}^2}{m_a^2})\cr
&\cdot &\sum_{kjm} {\rm Im}[(Y_\nu^{ki})^*Y_\nu^{kj} m_{\ell j}^2
(Y_\nu^{mj})^*Y_\nu^{mi}] \left( {\rm Log}\frac{M_{GUT}^2}{M_k^2}
\right)^2 {\rm Log}{M_{GUT}^2 \over M_m^2} \vec{S},\label{demu}\ea
where \ba h(x_L,x_E)& =&-{1 \over (x_E-x_L)^3}\left({1-x_L^2+2 x_L
{\rm Log} x_L \over (1-x_L)^3}-{1-x_E^2+2 x_E {\rm Log} x_E \over
(1-x_E)^3}\right)\cr &-&{1 \over
2(x_E-x_L)^2}\left({5-4x_L-x_L^2+2(1+2x_L){\rm Log}x_L \over
(1-x_L)^4} \right. \cr &&\left. +{5-4x_E-x_E^2+2(1+2x_E){\rm
Log}x_E \over (1-x_E)^4} \right).\ea
 Note that  one should insert the value of $\mu$ at the
supersymmetry breaking scale in Eq. (\ref{demu}).

To evaluate the order of magnitude of the EDMs, at first sight it
seems that we can simply set all the supersymmetric parameters to
some common scale $m_{susy}$ and take the values of the functions
$f$ and $h$ in Eqs. (\ref{dea0},\ref{dem1bar2},\ref{demu}) to be
numbers of order 1. However, this is not a valid simplification
because the functions $f$ and $h$ rapidly decrease when their
arguments fall below unity. In the  mSUGRA model we expect the
mass of the lightest neutralino to be smaller than that of
sfermions. As a result, we expect $h$ and $g$ to be smaller than
one. In section 4, we will see that this effect gives rise to a
suppression by two to three orders of magnitude.

\section{Effects of the phases of $\mu$ and $a_0$ on EDMs}
In this section, we first review the current bounds on $d_e$,
$d_\mu$, $d_D$, $d_{Hg}$ and $d_n$ and the prospects of improving
them. We then review how we can write them in terms of Im$(\mu)$
and Im$(a_0)$.
\begin{itemize}
\item {\bf Electron EDM $d_e$:}
The present bound on the EDM of electron is \be d_e<1.7\times
10^{-27} \ \ e~{\rm cm} \, \ \  \mbox{at}\, \ \ 95~\%\  {\rm CL}\
\cite{pdg} \ee DeMille and his Yale group are running an
experiment that uses the PbO molecules to probe $d_e$. Within
three years they can reach a sensitivity of $10^{-29}$~e~cm
\cite{yale}  and hopefully down to a sensitivity of
$10^{-31}$~e~cm within five years. There are proposals
\cite{solid} for probing $d_e$ down to $10^{-35}$~e~cm level. In
sum there is a very good prospect of measuring $d_e$ in future
\cite{4445}.
\item {\bf Neutron EDM $d_n$:} The present bound on
$d_n$ \cite{harris} is \be d_n<6.3\times 10^{-26} e~cm \ \ {\rm at
}\ \ 90 \% \ \ {\rm C.L.}\ee This bound will be improved
considerably by LANSCE \cite{lan} which will be able to probe
$d_n$ down to $4 \times 10^{-28}$~e~cm.
\item {\bf Muon EDM, $d_\mu$:}   The present bound on
$d_\mu$ \cite{pdg} is \be d_\mu<7\times 10^{-19} ~e~cm.\ee There
are proposals to measure $d_\mu$ down to $10^{-24}$~e~cm
\cite{futuremuon}. Using the storage ring of a neutrino factory,
measurement of $d_\mu$ down to $5\times 10^{-26}$ will become a
possibility \cite{nufactory}.

\item {\bf Mercury EDM $d_{Hg}$:}
The present bound on $d_{Hg}$ is \be \label{mercury} |d_{Hg}|<2.1
\times 10^{-28} e~cm \ee which can be improved by a factor of four
\cite{hgexp}.

\item {\bf Deuteron EDM $d_D$:}  The present bound on $d_D$ is very weak; however, there are
proposals \cite{Dexp} to probe $d_D$ down to \be |d_D|<(1-3)\times
10^{-27}~ e~cm \label{semer}. \ee

\end{itemize}
Different sources of CP-violation affect the EDMs listed above
differently. As a result, in principle by combining the
information on these observables, we can discriminate between
different sources of CP-violation. However to perform such an
analysis we must be able to express the EDMs in terms of
Im$[a_0]$, Im$[\mu]$ and Im$[Y_\nu]$. In the previous section, we
reviewed the effects of complex $Y_\nu$ on $d_e$. The effects of
complex $a_0$ and $\mu$ on $d_e$ are also well understood.
However, writing $d_n$, $d_{Hg}$ and $d_D$ in terms of  the
sources of CP-violation is more complicated. To do so, we first
have to express  $d_n$, $d_{Hg}$ and $d_D$ in terms  of the EDMs
and CEDMs of light quarks (namely, $d_u$, $d_d$, $d_s$,
$\tilde{d}_u$, $\tilde{d}_d$ and $\tilde{d}_s$) and then calculate
the quark EDMs and CEDMs in terms of Im$[a_0]$, Im$[\mu]$ and
Im$[Y_\nu]$. The quark EDMs and CEDMs in terms of Im$[a_0]$ and
Im$[\mu]$ have already been calculated in the literature. In this
paper we have used the results of Ref.  \cite{N=1}. As we
discussed in the appendix, the effects of Im$[Y_\nu]$ on the quark
EDMs and CEDMs are negligible. Unfortunately, the first step
(expressing $d_n$, $d_{Hg}$ and $d_D$ in terms of the quark EDMs
and CEDMs) is quite challenging. Let us consider them one by one.
\begin{itemize}
\item {\bf $d_n(d_q,\tilde{d}_q)$:}
\newline
Despite of the  rich literature on $d_n$ in terms of the quark
EDMs and CEDMs, the results are quite model dependent. For
example, the SU(3) chiral model \cite{su(3)chiral} and QCD sum
rules \cite{sum-rules} predict different contributions from
$\tilde{d}_u$ and $\tilde{d}_d$ to $d_n$. Considering these
discrepancies in the literature, in this paper we do not use
bounds on $d_n$ in our analysis. As it is shown in \cite{falk},
information on $d_n$ can help to refute the ``cancelation"
scenario. We will come back to this point later.

\item {\bf $d_{Hg}(d_q,\tilde{d}_q)$:}
\newline
There is an extensive literature on $d_{Hg}$ \cite{dhg}. In this
paper, following Ref. \cite{210-26}, we will interpret the bound
on $d_{Hg}$ as \be |\tilde{d}_d-\tilde{d}_u|<2 \times 10^{-26}~
{\rm cm}.\ee As shown in the recent paper \cite{adam}, the EDM of
electrons in the mercury atom can give a non-negligible
contribution to $d_{Hg}$. As a result, improvements on the bound
on $d_{Hg}$ will not be very helpful for us to discriminate
between different sources of CP-violation; {\it i. e.,} $d_{Hg}$
also obtains a correction from complex $Y_\nu$ through $d_e$.

\item {\bf $d_D(d_q,\tilde{d}_q)$:}
\newline
Searches for  $d_D$ can serve as an ideal probe for the  existence
of sources of CP-violation other than complex $Y_\nu$ because $i)$
there is a good prospect of improving the bound on $d_D$
\cite{Dexp}; $ii)$  an ionized deuteron does not contain any
electrons and hence we expect only a negligible and undetectable
contribution from $Y_\nu$ to $d_D$.

To calculate $d_D$ in terms of quark EDMs and CEDMs, two techniques have been suggested in the literature: $i)$
QCD sum rules \cite{lebedev} and $ii)$  SU(3) chiral theory \cite{su(3)chiral}. Within the error bars, the two
models agree on the contribution from $\tilde{d}_d-\tilde{d}_u$ which is the dominant one. However, the results
of the two models on the sub-dominant contributions are not compatible. Apart from this discrepancy, there is a
large uncertainty in the contribution of the dominant term: \be d_D(d_q,\tilde{d}_q)\simeq
-e(\tilde{d}_u-\tilde{d}_d)\,5^{+11}_{-3} \ . \ee In this paper we take ``the best fit" for our analysis.
\end{itemize}

\section{Numerical analysis}
\begin{figure}
\psfig{figure=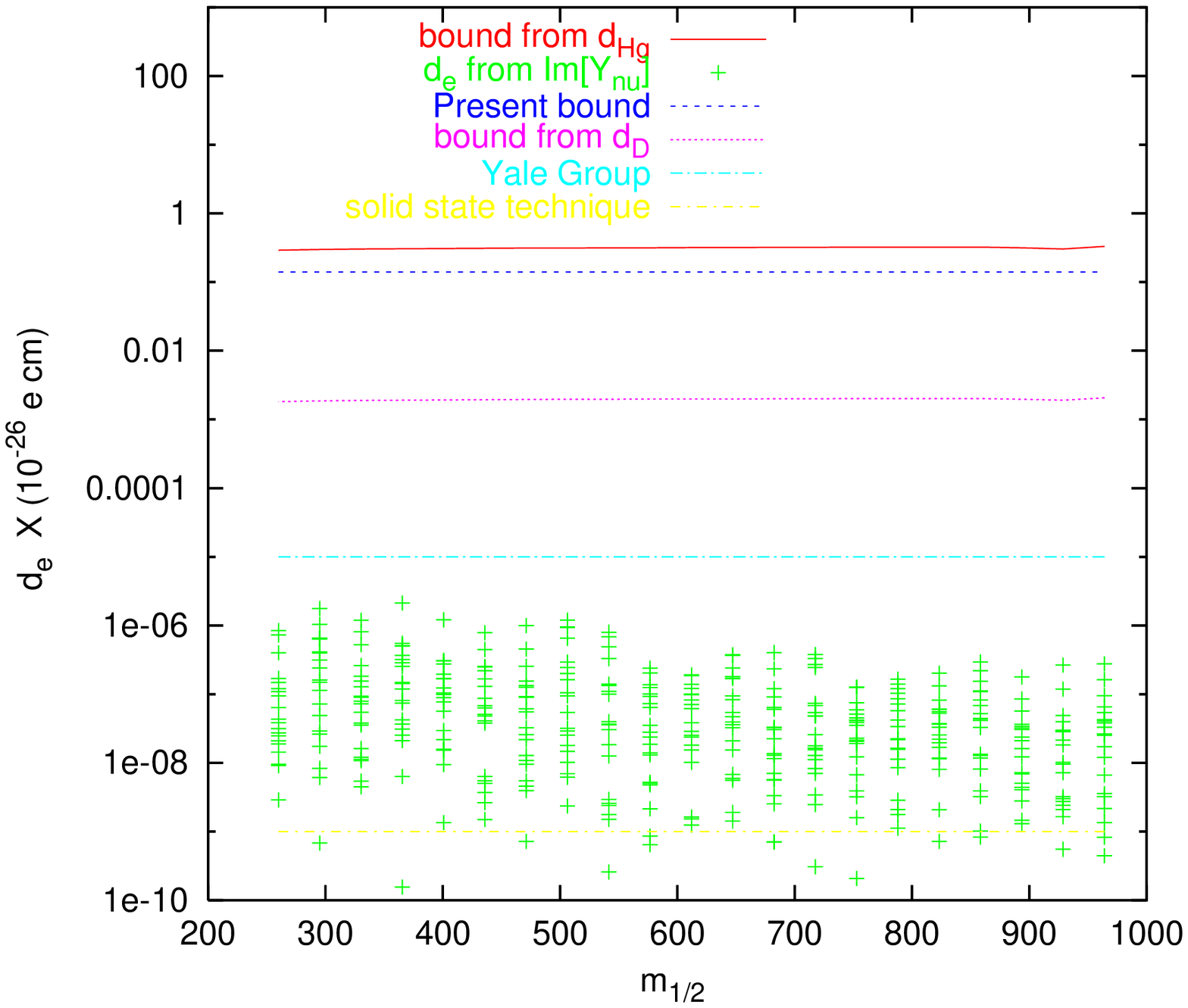,bb=50 28 545 477, clip=true,
 height=5 in
 } \caption{Electric dipole moment of the
electron for $a_0=0$, $\tan \beta=10$ and sgn($\mu)=+$. To draw
the red solid and purple dotted lines, we have assumed that
Im[$\mu$] is the only source of CP-violation and have taken
$\tilde{d}_d-\tilde{d}_u$ equal to  $2 \times 10^{-26}$~cm and $2
\times 10^{-28}$~cm, respectively to derive  Im[$\mu$]. To produce
the dots, we have assumed that complex $Y_\nu$ is the only source
of CP-violation and have randomly produced $Y_\nu$ compatible with
the data. The blue dashed line is the present bound on $d_e$
\cite{pdg} and  dot-dashed lines show the values of $d_e$ that can
be probed in the future \cite{yale,solid}}
 \label{a00tan10}
\end{figure}

\begin{figure}
\psfig{figure=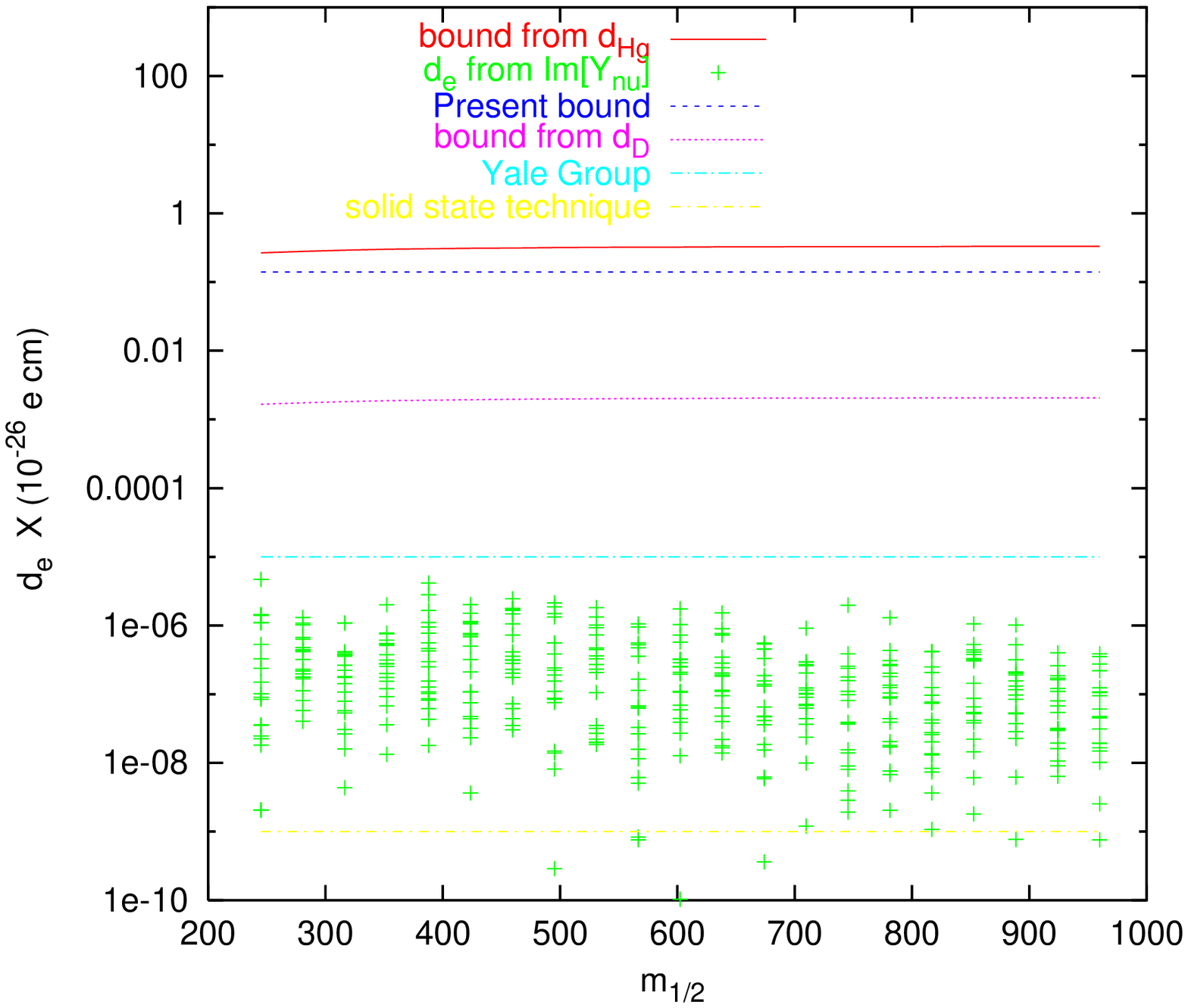,bb=50 28 545 477, clip=true,
height=5 in} \caption{The same as fig. \ref{a00tan10} for $a_0=0$,
$\tan\beta=20$ and sgn($\mu)=+$.} \label{a00tan20}
\end{figure}

\begin{figure}
\psfig{figure=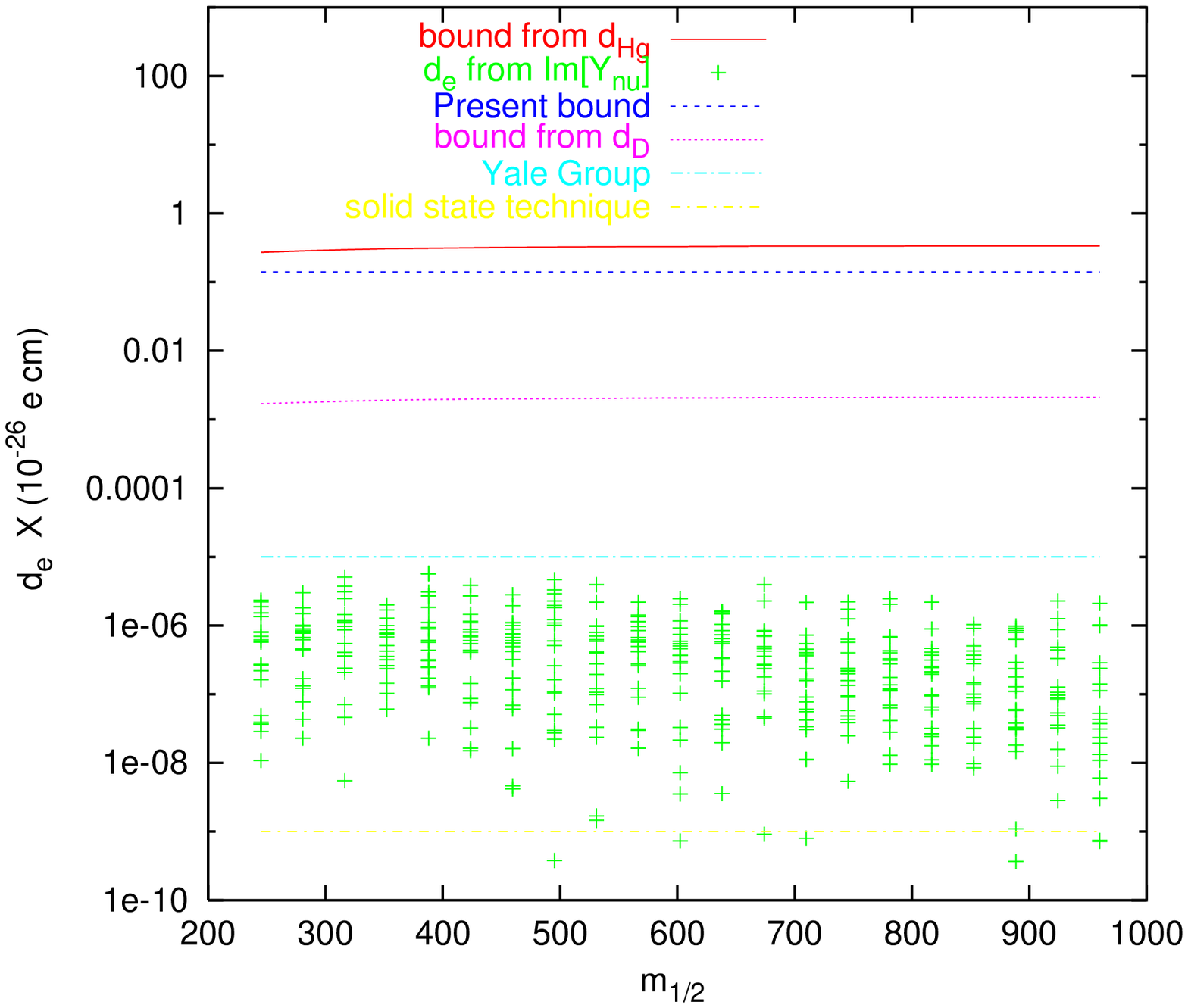,bb=50 28 545 477, clip=true,
height=5 in} \caption{The same as fig. \ref{a00tan10} for $a_0=0$,
$\tan\beta=20$ and sgn($\mu)=-$.} \label{a00tan20Minus}
\end{figure}

\begin{figure}
\psfig{figure=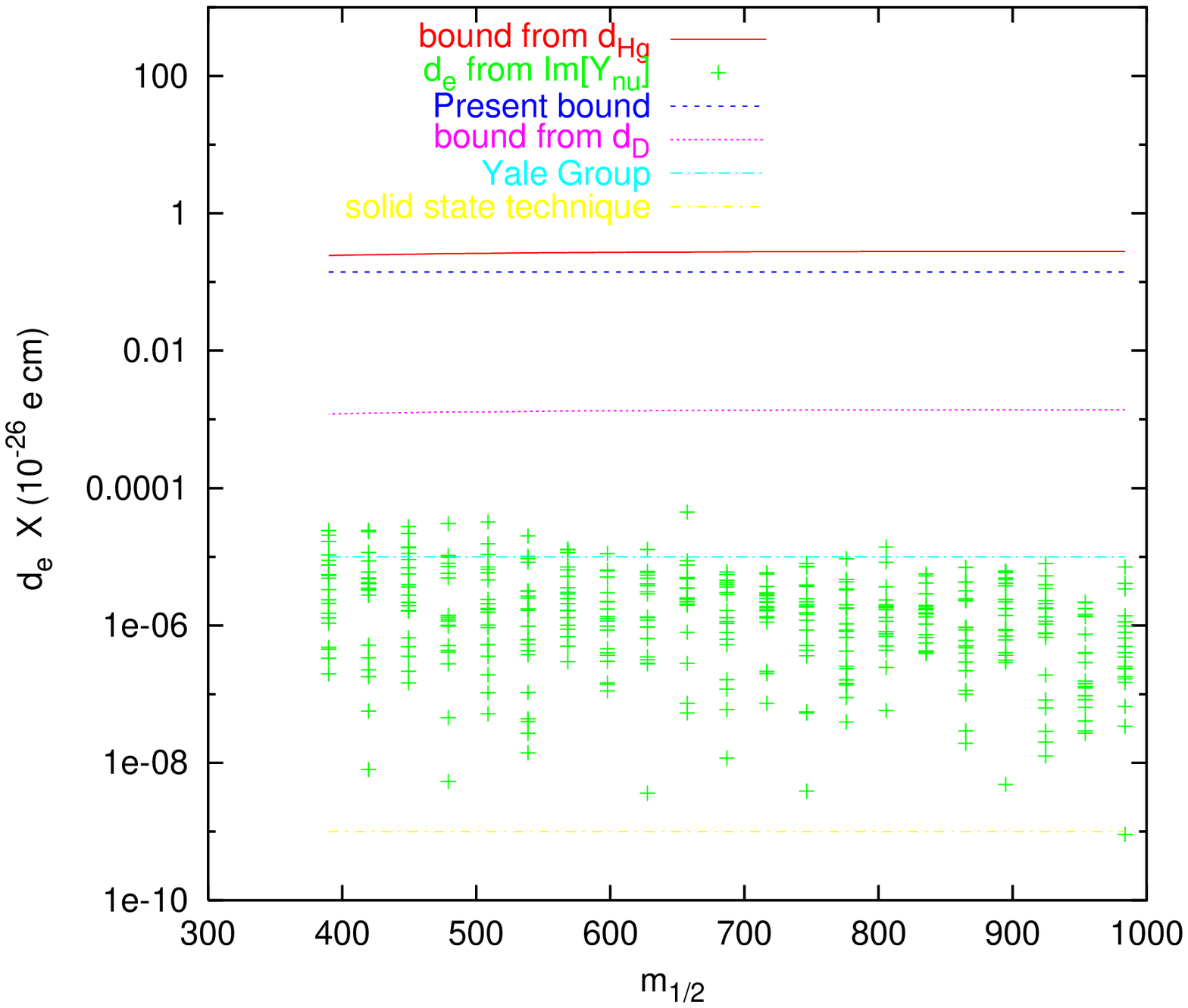,bb=50 28 545 500, clip=true,
height=5 in} \caption{Electric dipole moment of the electron for
$a_0=1000$~GeV, $\tan \beta=10$ and sgn($\mu)=+$. To draw the red
solid and purple dotted lines, we have assumed that Im[$a_0$] is
the only source of CP-violation and have taken
$\tilde{d}_d-\tilde{d}_u$ equal to  $2 \times 10^{-26}$~cm and $2
\times 10^{-28}$~cm, respectively to derive  Im[$a_0$]. To produce
the dots, we have assumed that complex $Y_\nu$ is the only source
of CP-violation and have randomly produced $Y_\nu$ compatible with
the data. The blue dashed line is the present bound on $d_e$
\cite{pdg} and  dot-dashed lines show the values of $d_e$ that can
be probed in the future \cite{yale,solid}}
 \label{a01000tan10}
\end{figure}

\begin{figure}
\psfig{figure=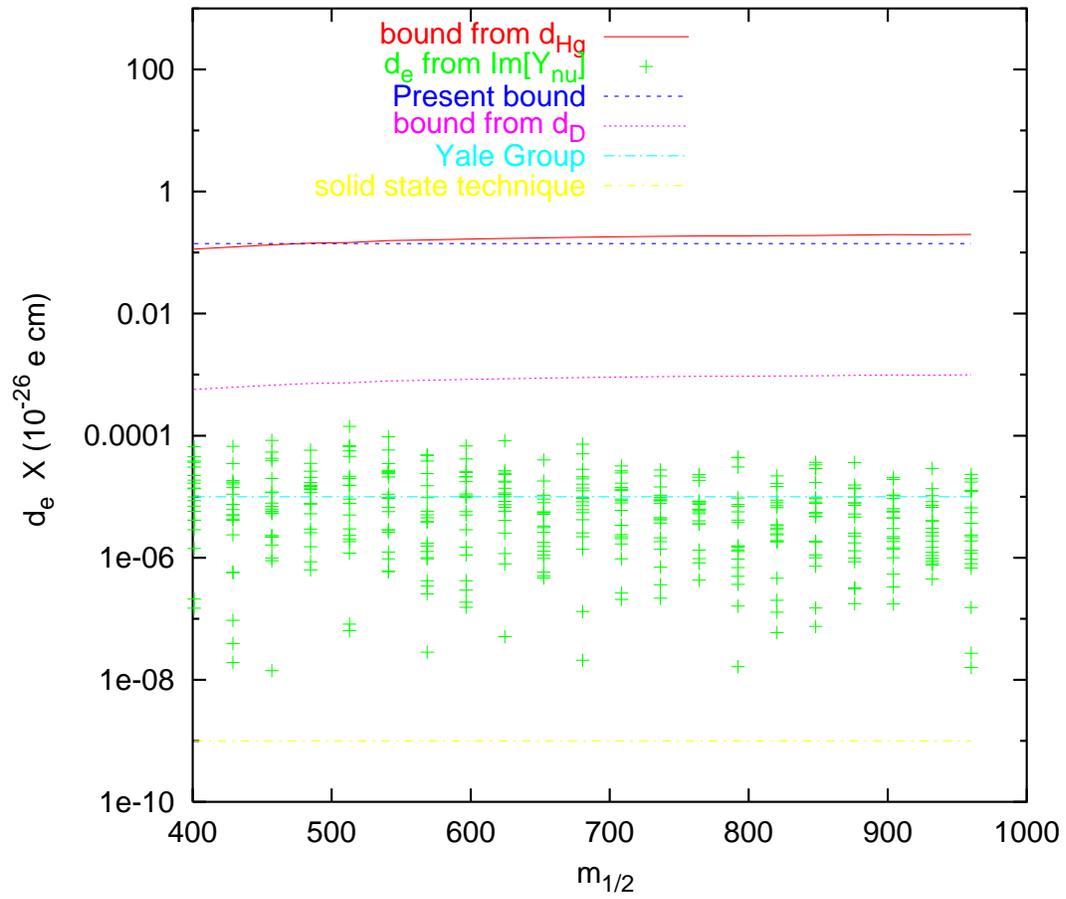,bb=50 28 545 477, clip=true,
height=5 in} \caption{The same as fig. \ref{a01000tan10} for
$a_0=1000$~GeV, $\tan \beta=20$ and sgn($\mu)=+$.}
\label{a01000tan20}
\end{figure}

\begin{figure}
\psfig{figure=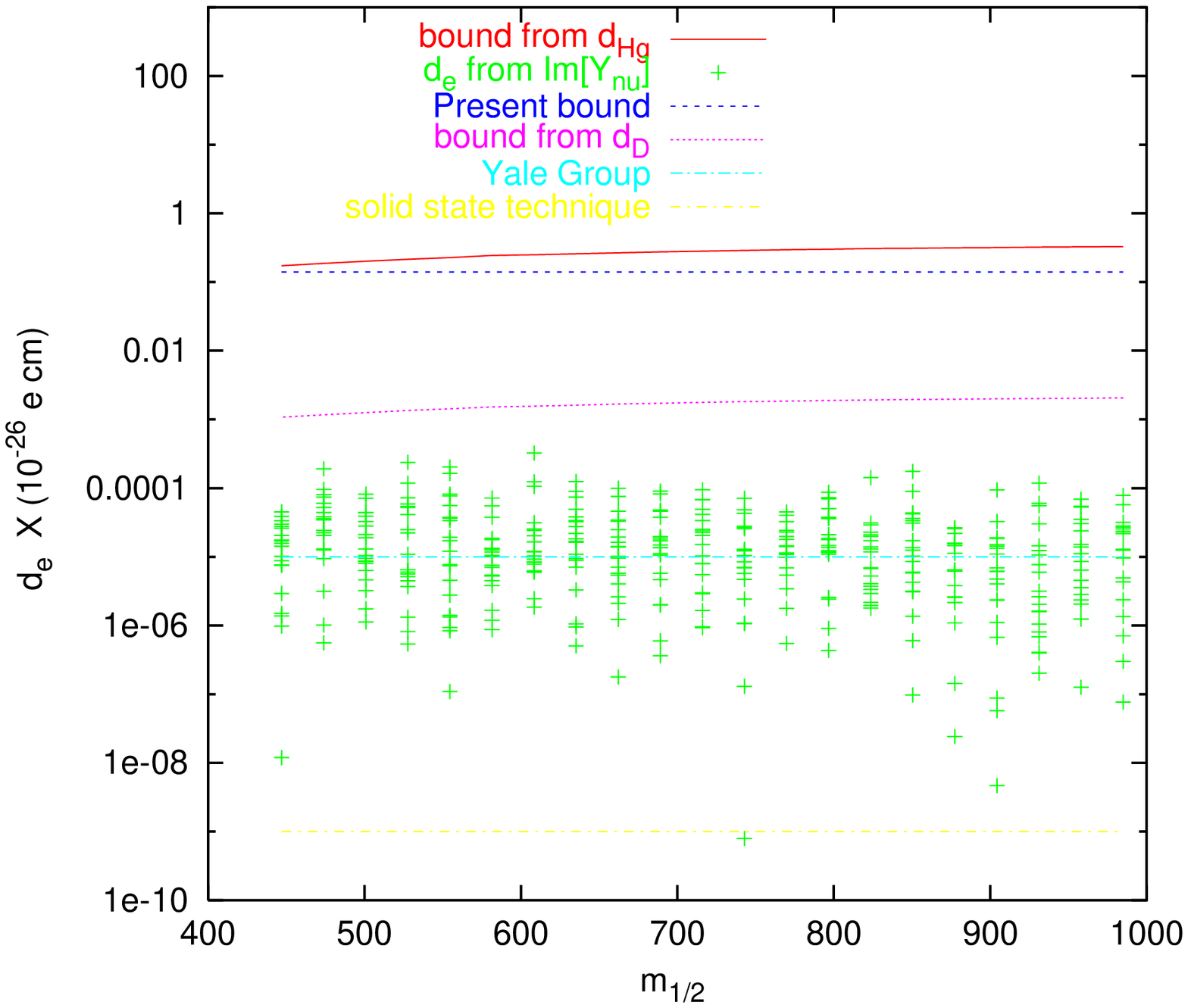,bb=50 28 545 477, clip=true,
 height=5 in
 } \caption{Electric dipole moment of the
electron for $a_0=2000$~GeV, $\tan \beta=20$ and sgn($\mu)=+$. To
draw the red solid and purple dotted lines, we have assumed that
Im[$\mu$] is the only source of CP-violation and have taken
$\tilde{d}_d-\tilde{d}_u$ equal to  $2 \times 10^{-26}$~cm and $2
\times 10^{-28}$~cm, respectively to derive  Im[$\mu$]. To produce
the dots, we have assumed that complex $Y_\nu$ is the only source
of CP-violation and have randomly produced $Y_\nu$ compatible with
the data. The blue dashed line is the present bound on $d_e$
\cite{pdg} and  dot-dashed lines show the values of $d_e$ that can
be probed in the future \cite{yale,solid}}
 \label{a02000tan20}
\end{figure}

In this section, we first describe how we  produce the random
seesaw parameters compatible with the data. We then describe the
figs. (\ref{a00tan10}-\ref{a02000tan20}) and, in the end, discuss
what can be inferred from the future data considering different
possible situations one by one.

In figures (\ref{a00tan10}-\ref{a02000tan20}), the dots marked
with "+" represent $d_e$ resulting from complex $Y_\nu$. To
extract random $Y_\nu$ and $M_N$ compatible with data, we have
followed the recipe described in \cite{recipe} and solved the
following two equations \be \eta_{m_\nu} Y_\nu^T {1\over M}Y_\nu
(v^2\sin^2 \beta) /2=U\cdot \Phi \cdot M_\nu^{Diag} \cdot \Phi
\cdot U^T \ee and \ba h\equiv Y_\nu^\dagger {\rm Log}{ M_{GUT}
\over M} Y_\nu=\left[ \matrix{ a &0 & d \cr 0& b& 0 \cr d^* & 0&
c}\right], \label{h}\ea where $v=247$ GeV, $M$ is the mass matrix
of the right-handed neutrinos, $U$ is the mixing matrix of
neutrinos with $s_{13}=0$ and $\Phi$ is $diag[1,e^{i\phi_1},e^{i
\phi_2}]$ with random values of $\phi_1$ and $\phi_2$ in the range
$(0,2\pi)$. Finally, $M_\nu^{Diag}= diag[m_1,\sqrt{m_1^2+\Delta
m_{21}^2},\sqrt{m_1^2+\Delta m_{31}^2}]$ where $m_1$ picks up
random values between 0 and 0.5 eV in a linear scale. The upper
limit on $m_1$   is what has been found in \cite{hannestad} by
taking the dark energy equation of state a free (but constant)
parameter. In the above equation, $\eta_{m_\nu}$ takes care of the
running of the neutrino mass matrix from $M$ to $M_{susy}$. Since
the deviation of $\eta_{m_\nu}$ from unity is small
\cite{running}, we have set $\eta_{m_\nu}=1$.

In order to satisfy the strong bounds on $\mbox{Br}(\mu \to e
\gamma$) \cite{pdg} and $\mbox{Br}(\tau \to \mu \gamma )$
\cite{newboundontaumu}, the matrix $h$, defined in Eq. (\ref{h}),
is taken to have this specific pattern with zero $e \mu$ and $\mu
\tau$ elements. Actually these branching ratios put bounds on
$(\Delta m_{\tilde{L}}^2)_{e \mu}$ and $(\Delta
m_{\tilde{L}}^2)_{\mu \tau}$ rather than on $h_{e \mu}$ and
$h_{\mu \tau}$. Notice that only the dominant term of $\Delta
m_{\tilde{L}}^2$ is proportional to $h$. There is also a
subdominant ``finite" contribution to $\Delta m_{\tilde{L}}^2$
which is about 10\% of the dominant effect and is not proportional
to the matrix $h$ \cite{peskin}. As we saw in section 2, this
finite part plays a crucial role in giving rise to EDMs because
the dominant leading-log part cancels out. Nonetheless, for
extracting the seesaw parameters, 20\% accuracy is enough and we
can neglect the subdominant part and take $\Delta m_{\tilde{L}}^2$
proportional to the matrix $h$. In Eq. (\ref{h}), $a$, $b$, $c$
are real numbers which take random
 values
  between 0 and 5. On the other hand, $|d|$ takes random values between 0 and the
 upper bound from $\mbox{Br}(\tau \to e \gamma)$ \cite{newboundontaue}. To calculate
 the upper  bound on $|d|$, we have used the formulae derived in
Ref. \cite{carlos}. The phase of $d$  takes random values between
0 and 2$\pi$. With the above bounds  on the random variables, the
Yukawa couplings can be relatively large, giving rise to \be {\rm
Im}\left[Y_\nu^\dagger {\rm Log}^2{M_{GUT}^2\over M_N^2} Y_\nu
Y_\nu^\dagger {\rm Log}{M_{GUT}^2\over M_N^2}
Y_\nu\right]_{ee}\sim {\rm few}\times 10\ee and \be {\rm
Im}\left[Y_\nu^\dagger {\rm Log}{M_{GUT}^2\over M_N^2} Y_\nu
Y_\nu^\dagger  Y_\nu\right]_{ee}\sim {\rm few}\times 0.1.\ee

As we discussed in the end of Sec. 2, because of the presence of
the rapidly changing functions $g(x_L,x_R)$ and $h(x_L,x_R)$ in
Eqs. (\ref{dea0}, \ref{dem1bar2}, \ref{demu}), the value of $d_e$
strongly depends on the values of the supersymmetric parameters.
To perform this analysis we have taken various values of $\tan
\beta$ and $a_0$ and calculated the spectrum of the supersymmetric
parameters along the $m_{1/2}-m_0$ strips parameterized in Ref.
\cite{petra}. Notice that Ref. \cite{petra} has already removed
the parameter range for which color or charge condensation takes
place.

In the figures, we have also drawn the present bound on $d_e$
\cite{pdg} as well as the limits  which can be probed in the
future. The present bound is shown by a dashed dark blue line and
lies several orders of magnitude above the $d_e$ from phases of
$Y_\nu$. After five years of data-taking, the Yale group can probe
$d_e$ down to $10^{-31}$ {\it e}~cm \cite{yale} which is shown
with a dot-dashed cyan line in the figures. As it is demonstrated
in the figures, only  for large values of $a_0$ the effect of
complex $Y_\nu$ on $d_e$ can be probed by the Yale group and for
most of the parameter space the effect remains beyond the reach of
this experiment.

There are proposals \cite{solid} to use solid state techniques to
probe $d_e$ down to $10^{-35}$ {\it e}~cm (shown with dot-dashed
yellow line in the figure). In this case, as it can be deduced
from the figure, we will have a great chance of being sensitive to
the effects of the phases of $Y_\nu$ on $d_e$. However,
unfortunately, the feasibility and time scale of the solid state
technique is still uncertain.

 Although for intermediate values of $\tan \beta$, the effect of the phases of $Y_\nu$ on $d_e$
is very low ($<10^{-30}$ {\it e}~cm ), its effect can still be
much higher than the four-loop effect on $d_e$ in the SM (the
effect of the CP-violating phase of the CKM matrix) which is
estimated to be $\sim 10^{-38}$ {\it e}~cm  \cite{ckm}.

In figs. (\ref{a00tan10}-\ref{a00tan20Minus}) as well as in fig.
\ref{a02000tan20}, $d_e$ resulting from Im[$\mu$] is also
depicted. The red solid lines in these figures show $d_e$ from
Im[$\mu$] assuming that the corresponding $d_{Hg}$ saturates the
present bound  \cite{hgexp}. As it is well-known, there are
uncertainties both in the value of $m_d$ \cite{pdg} and in the
interpretation of $d_{Hg}$ in terms of more fundamental parameters
$\tilde{d}_u$, $\tilde{d}_d$ and $\tilde{d}_s$. To draw this curve
we have assumed $m_d=5$ MeV and $\tilde{d}_u-\tilde{d}_d<2 \times
10^{-26}$ {\it e}~cm . As it is shown in the figure this bound is
weaker than even the present direct bound on $d_e$.
 The purple dotted lines in figs. (\ref{a00tan10}, \ref{a00tan20}, \ref{a00tan20Minus}, \ref{a02000tan20}),
 represent $d_e$ induced by values of Im$[\mu]$ that give rise to
$\tilde{d}_u-\tilde{d}_d=2\times 10^{-28}$ cm (corresponding to
$d_{D}=10^{-27}$ {\it e}~cm  and $d_D= 5e(\tilde{d}_d-\tilde{d}_u)
$). Notice that these  curves lie  well below the direct bound on
$d_e$ but the Yale group will be able to probe even smaller values
of $d_e$.
Similarly in figs. (\ref{a01000tan10},\ref{a01000tan20}), $d_e$
resulting from Im[$a_0$] is depicted.

 The following comments are in order:

1) The combination of the seesaw parameters that enter the formula
for $d_e$ resulting from Im$[\Delta m_{\tilde{E}}^2 m_\ell^2
\Delta m_{\tilde{L}}^2]$ [see Eq. (\ref{demu})] is \be {\rm
Im}\left[ Y_\nu^\dagger {\rm Ln}^2{M_{GUT}^2 \over M^2} Y_\nu
m_\ell^2Y_\nu^\dagger {\rm Ln}{M_{GUT}^2 \over M^2} Y_\nu
\right]_{ee}\simeq m_\tau^2{\rm Im}\left[ \left( Y_\nu^\dagger
{\rm Ln}^2{M_{GUT}^2 \over M^2} Y_\nu \right)_{e \tau}
\left(Y_\nu^\dagger {\rm Ln}{M_{GUT}^2 \over M^2} Y_\nu
\right)_{\tau e} \right] \ee where $M$  is the mass matrix of the
right-handed neutrinos. In contrast to this, the ``new" effect
given in Eq. (\ref{dem1bar2}) is proportional to  $${\rm Im}\left[
Y_\nu^\dagger {\rm Ln}^2{M_{GUT}^2 \over M_N^2} Y_\nu
Y_\nu^\dagger {\rm Ln}{M_{GUT}^2 \over M_N^2} Y_\nu \right]_{ee}.
$$ For the specific pattern of the $h$ matrix shown in Eq.
(\ref{h}) (with zero $e \mu$ element)  this effect is also given
by \be{\rm Im}\left[ \left( Y_\nu^\dagger {\rm Ln}^2{M_{GUT}^2
\over M_N^2} Y_\nu \right)_{e \tau} \left(Y_\nu^\dagger {\rm
Ln}{M_{GUT}^2 \over M_N^2} Y_\nu \right)_{\tau e} \right]. \ee In
other words, the two effects are proportional to each other.

For  the values of supersymmetric parameters chosen in Fig.
(\ref{a00tan10}) (that is, sgn($\mu$)=+, $\tan\beta=10$, $a_0=0$),
the "new" effect is dominant and is $-5$ times the effect
previously discussed in the literature. However, for
$a_0=1000$~GeV and 2000~GeV (Figs.
\ref{a01000tan10},\ref{a01000tan20} and \ref{a02000tan20}) the
dominant contribution is the one given by Eq. (\ref{demu}).

 2) In the figures, the bounds from $d_{Hg}$ and $d_D$ appear
 almost as horizontal lines. This results from the fact that for the $m_0-m_{1/2}$
 strips that
 we analyze, $m_0$ is almost proportional to $m_{1/2}$. Using dimensional analysis
 we can write
 $$ \tilde{d}_u-\tilde{d}_d \simeq k_1 {{\rm Im}[\mu] \ \ {\rm or \ \ Im}[a_0] \over
 m_{1/2}^3} \ \ \ \ \ \ d_e \simeq k_2 {{\rm Im}[\mu]  \ \ {\rm or \ \ Im}[a_0]\over m_{1/2}^3 }$$
 where $k_1$ and $k_2$ are given by the relevant fermion masses
 and are independent of $m_{1/2}$.
 As a result, for a given value of $\tilde{d}_u-\tilde{d}_d $,
 Im[$\mu$] (or Im[$a_0$]) itself is proportional to $m_{1/2}^3$ so
 $d_e$ will not vary with $m_{1/2}$.

 3) As discussed in Ref. \cite{Olive}, at two-loop level,
the imaginary $a_0$ can induce an imaginary correction to the
Wino mass, giving rise to another contribution to the EDMs. In our
analysis, we have taken this effect into account but it seems to
be subdominant.

In the following, we will discuss what can be
inferred about the sources of CP-violation
from $d_e$ and $d_D$ if their values
(or the bounds on them) turn out to be in certain ranges.

According to the Figs. (\ref{a00tan10}-\ref{a00tan20Minus}), for
$a_0=0$, any signal found by the Yale group implies that there are
sources of CP-violation other than the phases of the Yukawa
couplings. However, for larger values of $a_0$, the effect of
$Y_\nu$ on the EDMs can be observed by the Yale group within five
years.  According to Figs. (\ref{a01000tan10}-\ref{a02000tan20}),
for $a_0 \gsim 1000$~GeV  EDMs originating from complex $Y_\nu$
can be large enough to be observed by the Yale group. Therefore,
if after five years the Yale group reports a null result, we can
derive bounds on certain combinations of  seesaw parameters and
$a_0$. At least it will be possible to discriminate between low
and high $a_0$ values. If after five years the Yale group reports
a null result, we can derive bounds on the seesaw parameters.
However, if the Yale group finds that  $ 10^{-31}~e~{\rm
cm}<d_e<10^{-29}$~{\it e}~cm  we will not be able to determine
whether $d_e$ originates from complex $Y_\nu$ or from more
familiar sources such as complex $a_0$ or $\mu$. To be able to
make such a distinction, values of $d_D$ down to
$10^{-28}-10^{-29}$ {\it e}~cm   have to be probed which, at the
moment, does not seem to be achievable.

 If future searches for $d_D$
find  $d_D>10^{-27}$ {\it e}~cm  but the Yale group finds $d_e<2
\times 10^{-29}$ {\it e}~cm  (this can be tested within only 3
years of data taking by the Yale group \cite{yale}), we might
conclude that the source of CP-violation is something other than
pure Im$[\mu]$ or Im$[a_0]$; {\it e.g.,} QCD $\theta$-term which
can give a significant contribution to $d_D$ but only a negligible
contribution to $d_e$. Another possibility is that there is a
cancelation between the contributions of Im$[\mu]$ and Im$[a_0]$
to $d_e$. The information on $d_n$ would then help us to resolve
this ambiguity provided that the theoretical uncertainties in
calculation of $d_n$ as well  as $d_D$ are sufficiently reduced.

On the other hand, if the Yale group detects $d_e>2\times
10^{-29}$ {\it e}~cm, we will expect that $d_D>10^{-27}$ {\it
e}~cm  which will be a strong motivation for building a deuteron
storage ring and searching for $d_D$. If such a detector finds a
null result, within this framework the explanation will be quite
non-trivial requiring some fine-tuned cancelation between
different contributions.

According to these figures, in the foreseeable future, we will not
be able to extract any information on the seesaw parameters from
EDMs, because even if we develop techniques to probe $d_e$ as
small as $10^{-35}\, {\rm e-cm}$, we will not be able to subtract
(or dismiss) the effect coming from Im$[\mu]$ and Im$[a_0]$ unless
we are able to probe $\tilde{d}_u-\tilde{d}_d$ at least 5 orders
of magnitude below its present bound which seems impractical.
Remember  that this is under the optimistic assumptions that the
mass of the lightest neutrino, $m_1$, and $\mbox{Br}(\tau \to e
\gamma)$ are close to their upper bounds and there is no
cancelation between different contributions to the EDMs.

  If, in the future, we realize that $m_1$ and $\mbox{Br}(\tau
\to e \gamma)$ are indeed close to the present upper bounds on
them and $a_0=0$ ($a_0=1000\ {\rm GeV}$)  but find
$d_e<10^{-35}${\it e}~cm ($d_e<10^{-34}$ {\it e}~cm ), we will be
able to draw bounds on the phases of $Y_\nu$ which along with the
information on the phases of the Dirac and Majorana phases of the
neutrino mass matrix and the CP-violating phase of the left-handed
slepton mass matrix may have some implication for leptogenesis.
This is however quite an unlikely situation.

Let us now discuss the $d_\mu$. As we saw in Sec. 2, the phases of
$Y_\nu$ manifest themselves in the $d_\mu$ through Im$[\Delta
A_\ell \Delta m_{\tilde{L}}^2]_{\mu \mu}$ and Im$[\Delta
m_{\tilde{E}}^2 m_\ell^2 \Delta m_{\tilde{L}}^2]_{\mu \mu}$. If
$a_0$ is a real number, the matrix $A_\ell$ remains Hermitian
\cite{peskin}. That is the radiative corrections due to $Y_\nu$
cannot induce nonzero
 Im$[\Delta A_\ell]_{ii}$. So, we can write $${\rm Im}[\Delta
A_\ell \Delta m_{\tilde{L}}^2]_{\mu \mu}={\rm Im}[(\Delta
A_\ell)_{\mu e} (\Delta m_{\tilde{L}}^2)_{e \mu}]+ {\rm Im}
[(\Delta A_\ell)_{\mu \tau}(\Delta m_{\tilde{L}}^2)_{\tau \mu}]
$$
and
$${\rm Im}[\Delta m_{\tilde{E}}^2 m_\ell^2\Delta
m_{\tilde{L}}^2]_{\mu \mu}={\rm Im}[(\Delta  m_{\tilde{E}}^2)_{\mu
e}m_e^2(\Delta m_{\tilde{L}}^2)_{e \mu}]+ {\rm Im} [(\Delta
m_{\tilde{E}}^2)_{\mu \tau} m_\tau^2(\Delta m_{\tilde{L}}^2)_{\tau
\mu}].$$ The strong bounds on $\mbox{Br}(\mu \to e \gamma$)
\cite{pdg} and $\mbox{Br}(\tau \to \mu \gamma$)
\cite{newboundontaumu} can be translated into bounds on $(\Delta
m_{\tilde{L}}^2)_{e \mu}$ and $(\Delta m_{\tilde{L}}^2)_{\tau
\mu}$ as well as the corresponding elements of $\Delta A_\ell$. As
a result, in the framework that imaginary $Y_\nu$ is the only
source of CP-violation, we expect
$$ d_\mu \sim d_e \frac{m_\mu}{m_e}
{(\Delta m_{\tilde{L}}^2)_{\tau \mu} \over (\Delta m_{\tilde{L}}^2)_{\tau e}}<10^{-31} ~e~{\rm cm},$$
which will not be observable even if the muon storage ring of a nu-factory is built \cite{nufactory}. On the
other hand, imaginary $a_0$ and $\mu$ induce $d_\mu \sim d_e m_\mu/m_e$ and allow $d_e$ to be as large as the
experimental upper  bound on it. In this case,
we may have a chance of observing $d_\mu$. Observing $d_\mu \gg
m_\mu d_e/ m_e$ will indicate  that this simplified version of the MSSM  is not valid.

\section{Summary and conclusions}
In this work we have studied EDMs of particles in the context of
supersymmetric seesaw mechanism. We have examined various
contributions to electron EDM  induced by the CP--odd phases in
the neutrino Yukawa matrix. Our analysis takes into account
various contributions available in the literature as well as a new
one, proportional to the gaugino masses, which is presented in
Eq.~(\ref{dem1bar2}).

In our discussions we have first produced random complex neutrino
Yukawa couplings consistent with the bounds from LFV rare decays
and then calculated the electron EDM they induce along post-WMAP
$m_0-m_{1/2}$ strips for given values of $\tan \beta$ and $a_0$
\cite{petra}. We have found that, for small values of $a_0$, the
new contribution (\ref{dem1bar2}) can be dominant over  the other
contributions from $Y_\nu$ that had already been studied in the
literature.

It turns out that for a realistic mass spectrum of supersymmetric
particles, there is an extra suppression factor of
$10^{-2}-10^{-3}$ with which we would not encounter if all the
supersymmetric masses were taken to be equal to each other. In
figs. \ref{a00tan10}-\ref{a02000tan20}, the values of $d_e$
corresponding to different random complex $Y_\nu$ textures are
represented by dots. For small values of $\tan \beta$ ($\tan
\beta<10$) and $a_0$ ($a_0<1000$~{\rm GeV}), $d_e$ induced by
$Y_\nu$ is beyond the reach of the ongoing experiments
\cite{yale}. Such values of $d_e$  can however be probed by the
proposed solid state  based  experiments \cite{solid}. For larger
values of $\tan \beta$ and/or $a_0$, the Yale group may be able to
detect the effects of complex $Y_\nu$ on $d_e$. As it is
demonstrated in Figs. \ref{a01000tan20} and \ref{a02000tan20}, for
$\tan \beta=20 $ and $a_0=1000-2000$ GeV, a large fraction of
parameter space yields $d_e$  detectable by the Yale group.
However, even in this case we will not be able to extract
information on the seesaw parameters from $d_e$ because the source
of CP-violation might be $a_0$ and/or $\mu$ rather than $Y_\nu$.
If the future searches for $d_D$ \cite{lebedev} find out that
$d_D>10^{-27}$ {\it e}~cm  then we will conclude that there is a
source of CP-violation other than complex $Y_\nu$. However, to
prove that the dominant contribution to $d_e$ detected by the Yale
group comes from complex $Y_\nu$-- hence to be able to extract
information on the seesaw parameters from it-- we should show that
$d_D<10^{-28}-10^{-29}$ {\it e}~cm which is beyond the reach of
even the current proposals. Notice that for the purpose of
discriminating between complex  $Y_\nu$ and $a_0/\mu$ as sources
of CP-violation, searching for $d_{Hg}$ is not very helpful
because mercury atom contains electron and hence $d_{Hg}$ obtains
a contribution from complex $Y_\nu$. That is while ionized
deuteron used for measuring $d_D$ does not contain any electron
and the contribution of complex $Y_\nu$ to it is negligible. To
obtain information from $d_n$, the theoretical uncertainties first
have to be resolved.

In this paper, we have also shown that for the neutrino Yukawa
couplings satisfying the current bounds from the LFV rare decays,
the electric dipole moment of muon induced by $Y_\nu$ is
negligible and cannot be detected in the foreseeable future.
Detecting a sizeable $d_\mu$ will indicate that there are sources
of CP-violation beyond the complex $Y_\nu$.

\section{Appendix} Since $d_D$ is dominantly given by
$\tilde{d}_d-\tilde{d}_u$, in this section, we concentrate on
evaluating  $\tilde{d}_q$. One can repeat a similar discussion for
$d_q$.

In Sec. 2, we saw that integrating out $N_i$, the effects of
CP-violating phases appear in the left-right mixing of sleptons
which can be evaluated to $m_\ell m_{susy} F(Y_\nu, {\rm Log}[
M_{GUT}^2/ M^2])/(16 \pi^2)^2$ or for  large  $\tan \beta$, to
$m_\ell m_{susy}\tan \beta (m_\tau^2 \tan^2\beta /v^2)
F'(Y_\nu,{\rm Log}[ M_{GUT}^2/ M^2])/(16 \pi^2)^2$. For random
$Y_\nu$ consistent with observed $m_\nu$ and bounds on the
branching ratios of LFV rare decay, the functions $F$ and $F'$
take values smaller than 0.1. Since quarks do not directly couple
to the leptonic sector, the CP-violation in the leptonic sector
should be transferred to the quark sector through one-loop (or
higher loop) effective operators made of Higgs and gauge bosons or
their superpartners. To construct  such an effective operator one
more factor of $Y_\ell$ is needed to compensate for the left-right
mixing mentioned above. Considering the fact that $Y_\tau\gg
Y_\mu, Y_e$, the main contribution to the effective operator comes
from the diagrams with $\tau$ and $\tilde{\tau}$ propagating in
them. So the CP-odd effective potential will be given by
$$m_{susy}^n m_\tau^2 \tan\beta F(Y_\nu, {\rm Log}[
M_{GUT}^2/ M^2])/(16 \pi^2)^3$$ or for large $\tan \beta$,
$$m_{susy}^n m_\tau^2 \tan^2\beta(m_\tau^2 \tan^2\beta /v^2) F'(Y_\nu,{\rm Log}[
M_{GUT}^2/ M^2])/(16 \pi^2)^3.$$ In these formula, the power of
$m_{susy}$, $n$, is determined by the dimension of the specific
operator under consideration.

To evaluate $\tilde{d}_q$, we have to insert the CP-odd effective
operator in another one-loop diagram. Since CEDMs mix left- and
right-handed, the latter diagram should involve a factor of $Y_q$.
So, we can write $$\tilde{d}_q\sim { Y_q g^2 g_s\over 16
\pi^2}{m_\tau^2 \tan\beta F(Y_\nu, {\rm Log}[ M_{GUT}^2/
M^2])\over (16 \pi^2)^3m_{susy}^3}$$ and for large $\tan \beta$,
 $$\tilde{d}_q\sim { Y_q g^2 g_s\over 16
\pi^2}{m_\tau^2 \tan^2\beta(m_\tau^2 \tan^2\beta /v^2)
F'(Y_\nu,{\rm Log}[ M_{GUT}^2/ M^2])\over (16
\pi^2)^3m_{susy}^3}.$$ As a result, we expect $\tilde{d}_d <
10^{-30} $~cm which cannot be observed even if the recent proposal
\cite{Dexp} is implemented. We expect $\tilde{d}_u$ to be even
smaller because $Y_u/Y_d=m_u/m_d\cot \beta\ll 1$. Notice that
although $\tilde{d}_q$ is suppressed by a factor of $m_\tau^2
\tan^2\beta/(16 \pi^2 m_{susy}^2)$, $e\tilde{d}_d$ can be
comparable to $d_e$. This originates from two facts: $Y_d/Y_e\sim
10$ and in the case of $d_e$, as we discussed in Sec. 2, there is
an extra suppression given by the functions $g(x_L,x_E)$ and
$h(x_L,x_E)$. If we do precise two-loop  calculation of
$\tilde{d}_d$ for a realistic SUSY spectrum, we may encounter
similar suppression. As the above analysis show we do not expect
an observable effect due to $Y_\nu$ in future searches for $d_D$
and $d_{Hg}$ so it seems  there is not a strong motivation for
performing such a complicated two-loop calculation.

\section{Acknowledgements}
The research of D.D. was partially supported by GEBIP grant of the
Turkish Academy of Sciences and by project 104T503 of Scientific
and Technical Research Council of Turkey. The authors are grateful
to J. Ellis and M. Peskin for the encouragement and useful
discussions. Y. F. would like to thank A. Ritz for fruitful
discussions. The authors also appreciate M. M. Sheikh-Jabbari for
careful reading of the manuscript, and for useful comments.

\newpage

\newpage


\end{document}